\documentclass[doublecol]{epl2}
\usepackage{graphicx,amsmath,amssymb,mathrsfs,eufrak}
\title{Tunable transmission due to defects in zigzag phosphorene nanoribbons}
\shorttitle{Tunable transmission due to defects in ...}
\author{Mohsen Amini, Morteza Soltani\footnote{mo.soltani@sci.ui.ac.ir}, Ebrahim Ghanbari-Adivi, and
Milad Sharbafiun} \shortauthor{M. Amini $et~al$}
\institute{ Department of Physics, Faculty of Sciences, University
of Isfahan, Isfahan 81746-73441, Iran}
\abstract{Transport of the edge-state electrons along zigzag
phosphorene nanoribbons in presence of two impurities/vacancies is
analytically investigated. Considering the places of the defects, a
number of different situations are examined. When both defects are
placed on the edge zigzag chain, as is expected, with changing the
energy of the traveling electrons the electrical conductance
exhibits a resonance behavior. In this case, for two vacancies the
observed resonant peaks become extremely sharp. An amazing behavior
is seen when the second vacancy is located along an armchair chain
while the first is placed at the intersection of the edge zigzag and
this armchair chains. In this case, in a considerable range of
energy, the conductance is strongly strengthened. In fact the
presence of the second vacancy create a shielded region around the
first vacancy, consequently, the traveling wave bypasses this region
and  enhances the conductivity. The analytical results are compared
with numerical simulations showing a very good agreement. }
\begin{document}
\maketitle
\section{Introduction\label{Sec1}}
Graphene, as the most fundamental two-dimensional~(2D) material in
the world, has no electronic band gap in its electronic structure
and this feature is an obstacle in practical application of this
material in the design of semiconductor devices~\cite{Neto2009}.
Therefore, it is natural to look for the applicable 2D~materials
with a gap in their band structures for such purposes. Phosphorene,
as a new elemental quasi-2D material, has recently attracted great
attention of the manufacturers of the electronic devices. This is
due to the large direct band gap of phosphorene ranged from about
0.5~eV~for the five-layer structures to 1.5~eV for the monolayer
ones~\cite{Qiao2014}.\par
%%%%%%%%%%%%%%%%%%%%%%%%%%%%%%%%%%%%%%%%%%%%%%%%%%%%%%%%%%%%%%%%%%%%%%%%%%%%%%%
Bulk phosphorus is a layered crystal of the van der Waals
type~\cite{Shulenburger2015} which can be exfoliated into
phosphorene as a separated single layer material~\cite{Liu2014}. The
peculiarity of phosphorene resides in the fact that each phosphorus
atom is bonded covalently with three nearest neighbors via $sp^{3}$
hybridization to make a puckered 2D honeycomb structure which gives
rise to various anisotropic
properties~\cite{Neto2014,Peeters2014,Guinea2014,Yang2014,Katsnelson2015}.
Furtheremore, zigzag phosphorene nanoribbon~(zPNR), which can be
fabricated in experiment~\cite{Das2016,Liang2014}, supports an edge
band in the middle of main gap which is completely isolated from
both the conduction and valence bands. Due to the presence of two
edge boundaries in zPNRs,  the quasi-flat band composed of edge
modes is doubly degenerated~\cite{Ezawa2014} with interesting
transport properties. Similar, edge states in skewed-armchair
phosphorene nanoribbons have been studied in Ref.~\cite{Grujic2016}.
In the absence of defect, both zigzag and skewed-armchair
phosphorene nanoribbons are metallic unless they undergo an edge
reconstruction or passivation~\cite{Carvalho2014}.\par
%%%%%%%%%%%%%%%%%%%%%%%%%%%%%%%%%%%%%%%%%%%%%%%%%%%%%%%%%%%%%%%%%%%%%%%%%%%%%%%
In addition to above mentioned properties,  the negative
differential resistance~(NDR) behavior which is robust with respect
to the edge reconstruction has been reported for a two-terminal zPNR
device~\cite{Maity2016}. Also, using the first-principles
calculations, it is shown  that a transverse electric field can
induce an insulator-metal transition in the zPNRs with
hydrogen-saturated edges~\cite{Feng2015}. Finally, the effect of
impurity defects on the quantum transport in zPNRs is investigated
recently~\cite{Li2018,Nourbakhsh2018,Amini2018}. The studies show
that the presence of impurity defects, ranging from
vacancies~\cite{Li2018} to substitution of atoms via
doping~\cite{Nourbakhsh2018}, changes the transport properties of
pristine zPNRs. The presence of localized  scattering centers near
the edges of zPNR, have a significant effect on the electronic
transmission through the edge states and reduces the transmittance
of the sample depending on its distance from the
edge\cite{Amini2018}. On the other hand, it is important to explore
new efficient ways to modulate the transport properties of zPNR such
as tuning the band structure by mechanical stress~\cite{Han2014}. A
possible subsequent step  in these studies is to investigate the
effect of the presence of multiple scattering centers, manipulated
on the atomic scale in a nanoribbon, on the quantum transport
properties of this material. For example, the influence of the edge
states on the Ruderman-Kittel-Kasuya-Yosida~(RKKY) exchange
interaction between two doped on-site magnetic impurities in a zPNR
has been recently investigated~\cite{Islam2018}.\par
%%%%%%%%%%%%%%%%%%%%%%%%%%%%%%%%%%%%%%%%%%%%%%%%%%%%%%%%%%%%%%%%%%%%%%%%%%%%%%%
In this paper, similar to what has been studied on the coherent
electron transport along zigzag graphene
nanoribbons\cite{Bahamon2010},  we study the effects of two
impurities/vacancies on the electron transport along a zigzag chain
in a zPNR. To this end, a Green's function approach, which we have
developed previously to study the influence of the impurity defects
on the scattering of the traveling electrons in a
zPNR~\cite{Amini2018,ASGS2018}, is employed. For different cases the
conductance is evaluated as a function of the electron energy. This
function displays a resonant structure when the defects are placed
along the edge zigzag chain. A more interesting behavior is observed
when the second vacancy is placed adjacent to the first one on an
armchair chain. This combination maximize the conduction in a
certain range of energy.\par
%%%%%%%%%%%%%%%%%%%%%%%%%%%%%%%%%%%%%%%%%%%%%%%%%%%%%%%%%%%%%%%%%%%%%%%%%%%%%%%
%%%%%%%%%%%%%%%%%%%%%%%%%%%%%%%%%%%%%%%%%%%%%%%%%%%%%%%%%%%%%%%%%%%%%%%%%%%%%%%
%%%%%%%%%%%%%%%%%%%%%%%%%%%%%%%%%%%%%%%%%%%%%%%%%%%%%%%%%%%%%%%%%%%%%%%%%%%%%%%
\begin{figure}[t]
\centering
\includegraphics[scale=0.55]{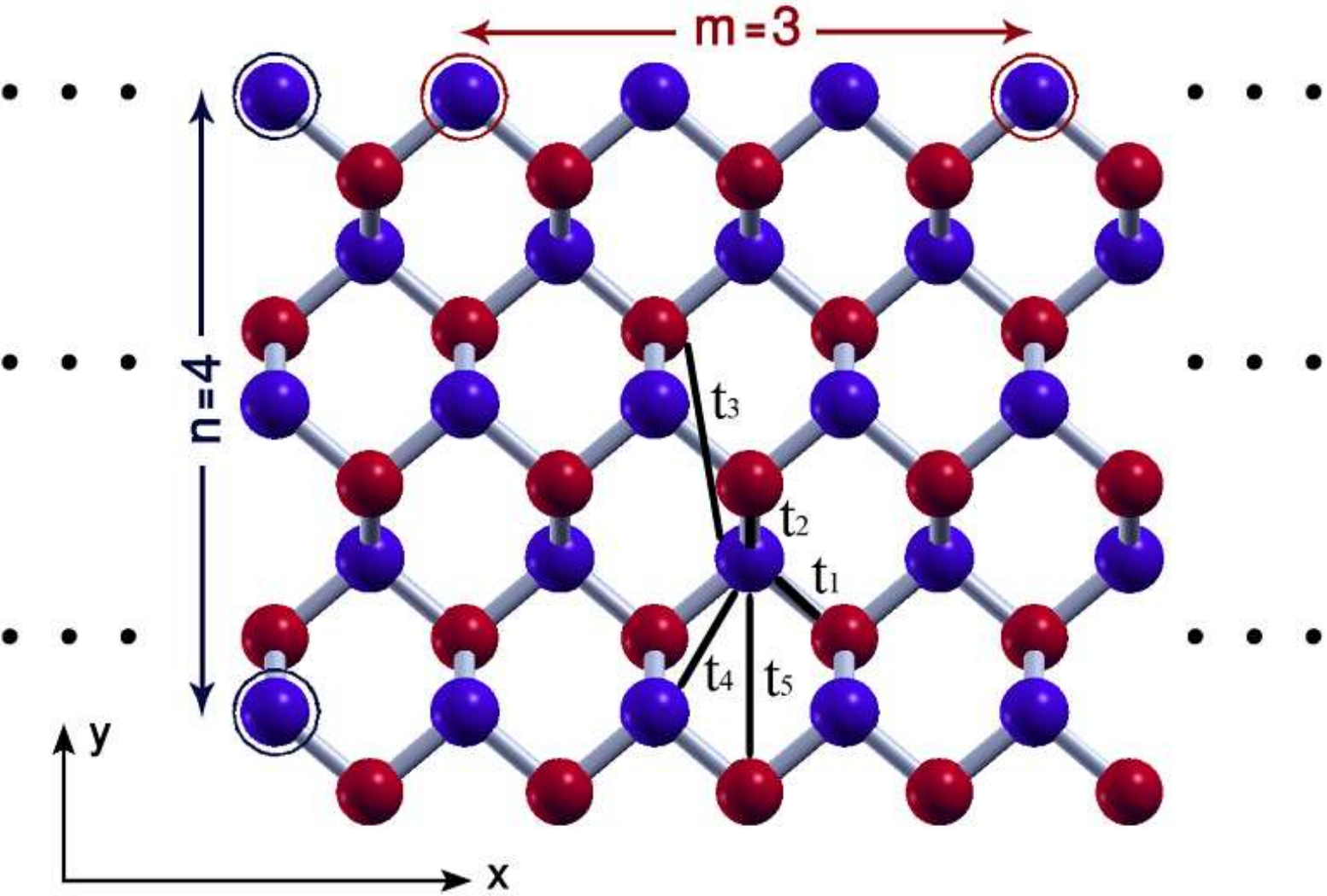}
\caption{(Color online) Schematic of the lattice structure of
phosphorene and hopping integrals $t_{i}$. The left and right edges
are armchair while the up and down edges are zigzag. Blue and red
balls represent phosphorus atoms in the lower and upper layer
respectively. Also, two-defect structures along the $x$ and $y$ axes
are shown schematically.} \label{Fig01}
\end{figure}
%%%%%%%%%%%%%%%%%%%%%%%%%%%%%%%%%%%%%%%%%%%%%%%%%%%%%%%%%%%%%%%%%%%%%%%%%%%%%%%
%%%%%%%%%%%%%%%%%%%%%%%%%%%%%%%%%%%%%%%%%%%%%%%%%%%%%%%%%%%%%%%%%%%%%%%%%%%%%%%
%%%%%%%%%%%%%%%%%%%%%%%%%%%%%%%%%%%%%%%%%%%%%%%%%%%%%%%%%%%%%%%%%%%%%%%%%%%%%%%
\section{Theory\label{Sec2}}
As is schematically shown in figure~\ref{Fig01}, phosphorene has a
non-planar puckered honeycomb lattice which consists of two
inequivalent sublattices denoted by $A$ and $B$. In a tight-binding
approach, the Hamiltonian of such a system including a number of the
on-site impurities is given as
%%%%%%%%%%%%%%%%%%%%%%%%%%%%%%%%%%%%%%%%%%%%%%%%%%%%%%%%%%%%%%%%%%%%%%%%%%%%%%%
\begin{equation}
\label{EQ01} H = \sum_{\langle i,j\rangle} t_{ij} c_{i}^\dagger c_j+
h.c.  + \hat{V},
\end{equation}
%%%%%%%%%%%%%%%%%%%%%%%%%%%%%%%%%%%%%%%%%%%%%%%%%%%%%%%%%%%%%%%%%%%%%%%%%%%%%%%
where $\langle i,j\rangle$ represents the index of summation that is
taken over only the considered nearest neighbors, $t_{ij}$ is
referred to as the hopping integral between atoms at sites $i$ and
$j$,  $c_i$ and $c^\dagger_i$ are the electron annihilation and
creation operators in site $i$,  $h.c.$ stands for Hermitian
conjugate and $V$ is the disorder potential energy. It is obvious
that in absence of the defects,  $V$ is zero.\par
%%%%%%%%%%%%%%%%%%%%%%%%%%%%%%%%%%%%%%%%%%%%%%%%%%%%%%%%%%%%%%%%%%%%%%%%%%%%%%%
The calculations based on the \emph{ab initio} tight-binding method,
show that taking only five nearest-neighbor interatomic interactions
is a very good approximation to provide a reasonable description of
the phosphorene band structure~\cite{Rudenko2014}. Referring to
Fig.~\ref{Fig01}, the hopping integrals corresponding to these
interactions are denoted for simplicity by $t_1$ to $t_5$ and their
corresponding values are assumed the same as reported in
Ref.~\cite{Rudenko2014}. These values are given as $t_1 =-
1.220~eV$, $t_2 = 3.665~eV $, $t_3 = - 0.205~eV$, $t_4 = -
0.105~eV$, and $t_5 = - 0.055~eV$. \par
%%%%%%%%%%%%%%%%%%%%%%%%%%%%%%%%%%%%%%%%%%%%%%%%%%%%%%%%%%%%%%%%%%%%%%%%%%%%%%%
Phosphorene can be cut into nanoribbons that the shape of their
edges depends on the direction of the cut. Two typical crystal
directions, namely the armchair and zigzag directions, were
generally explored. The nanoribbons with zigzag edges exhibit some
interesting behaviors representing important physical aspects.  One
of the interesting aspects of these nanoribbons is the existence of
quantum edge states in the edge zigzag chains. In the following
section, we are interested in studying the behavior of the
edge-state electrons, traveling along a side zigzag chain, when they
scatters off the impurities doped into the chain sites. In order to
investigate this issue, the electronic properties of this type of
the nanoribbons is studied analytically. Also, a numerical
simulation is made to study the tunable resonances due to the
impurities and the validity of the analytical calculations is
examined by comparing their obtained results with this numerical
simulation.\par
%%%%%%%%%%%%%%%%%%%%%%%%%%%%%%%%%%%%%%%%%%%%%%%%%%%%%%%%%%%%%%%%%%%%%%%%%%%%%%%
%%%%%%%%%%%%%%%%%%%%%%%%%%%%%%%%%%%%%%%%%%%%%%%%%%%%%%%%%%%%%%%%%%%%%%%%%%%%%%%
\begin{figure}[t]
\centering
\includegraphics[scale=0.8]{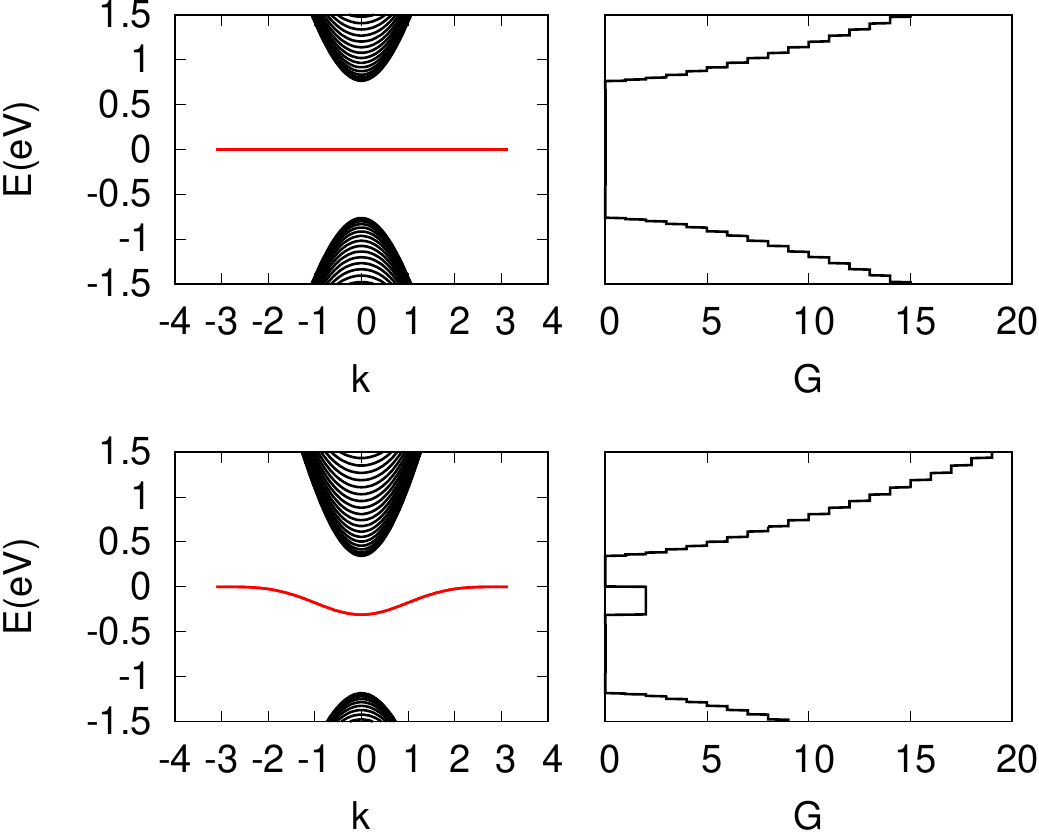}
\caption{(Color online) Conductance and band structure of zPNRs (a)
when the hopping integral $t_4$ is zero (b) when $t_4$ nonzero. The
flat band for the first case and the quasi-flat band for the second
case are marked in red.} \label{Fig02}
\end{figure}
%%%%%%%%%%%%%%%%%%%%%%%%%%%%%%%%%%%%%%%%%%%%%%%%%%%%%%%%%%%%%%%%%%%%%%%%%%%%%%%
%%%%%%%%%%%%%%%%%%%%%%%%%%%%%%%%%%%%%%%%%%%%%%%%%%%%%%%%%%%%%%%%%%%%%%%%%%%%%%%
%%%%%%%%%%%%%%%%%%%%%%%%%%%%%%%%%%%%%%%%%%%%%%%%%%%%%%%%%%%%%%%%%%%%%%%%%%%%%%%
%%%%%%%%%%%%%%%%%%%%%%%%%%%%%%%%%%%%%%%%%%%%%%%%%%%%%%%%%%%%%%%%%%%%%%%%%%%%%%%
%%%%%%%%%%%%%%%%%%%%%%%%%%%%%%%%%%%%%%%%%%%%%%%%%%%%%%%%%%%%%%%%%%%%%%%%%%%%%%%
\subsection{Quantum edge states in a zPNR}
As is known, a zigzag edge chain in a honeycomb lattice supports
some quantum edge states which are localized at the edge sites of
the side zigzag chains. The exitance of these edge states in zPNRs
has created a great potential for this material to use it in science
and technology. These states are also the essential ingredients in
many of the fascinating properties of phosphorene. The properties of
the edge states in phosphorene are intimately related to the
properties of the bulk band. The bulk energy gap can be opened by
breaking the inversion symmetry of the lattice. Both electron-hole
and the lattice inversion symmetries are broken in phosphorene by
the interatomic interaction corresponding to the hopping integral
$t_4$.\par
%%%%%%%%%%%%%%%%%%%%%%%%%%%%%%%%%%%%%%%%%%%%%%%%%%%%%%%%%%%%%%%%%%%%%%%%%%%%%%%
The first step in the  evaluation of the transport properties is to
calculate the conductance which depends on the Hamiltonian of the
specified system. Here, we use  the Landauer formula,
\begin{equation}
\label{EQ02} g(E) ={2 e^2\over h} \mathcal{T}(E),
\end{equation}
to evaluate the conductance of a zPNR. In this equation, $g(E)$ is
conductance, $e$ is the fundamental charge, $h$ is the Planck
constant and  $\mathcal{T}(E)$ is the transmission coefficient. Both
conductance $g(E)$ and the transmission coefficient $\mathcal{T}(E)$
depend on the electron energy $E$.\par
%%%%%%%%%%%%%%%%%%%%%%%%%%%%%%%%%%%%%%%%%%%%%%%%%%%%%%%%%%%%%%%%%%%%%%%%%%%%%%%
For two different cases, both conductance and band structure of a
typical zPNR are schematically shown in Fig.~\ref{Fig02}. For the
first case, we assumed that the hopping interaction $t_4$ is zero.
As is seen, in this case a flat edge state, which is completely
separated form both conduction and valence bands, appears in the
band structure. The second case is for a more real situation for
which $t_4$ is not zero. In this case the flat edge state tends to
become a quasi-flat state.\par
%%%%%%%%%%%%%%%%%%%%%%%%%%%%%%%%%%%%%%%%%%%%%%%%%%%%%%%%%%%%%%%%%%%%%%%%%%%%%%%
Employing the tight-binding hamiltonian given in Eq.~(\ref{EQ01})
and applying a first-order perturbation theory, a dispersion
relation can be derived for the edge states of zPNRs and the
properties represented in Fig.~\ref{Fig02} can be explained. To this
end, at first we neglect the interaction terms including the hopping
integrals of $t_3$ and $t_5$. Since $t_3$ and $t_5$ in comparison
with $t_1$ and $t_2$ have a smaller contribution to the Hamiltonian,
this assumption seems reasonable. Also, since $t_4$ is responsible
for the symmetries mentioned above, we consider the corresponding
interaction term in the hamiltonian as a perturbation.  Takeing $x$
and $y$ axes along the zigzag and armchair directions, each lattice
site can be labeled by a pair of integers $(m,n)$ where $m$ and $n$
are armchair and zigzag chain numbers. Without loss of generality,
we focuss on an A-type edge chain. For this case, we notice that a
typical unperturbed edge state is given as
%%%%%%%%%%%%%%%%%%%%%%%%%%%%%%%%%%%%%%%%%%%%%%%%%%%%%%%%%%%%%%%%%%%%%%%%%%%%%%%
\begin{equation}
\label{EQ03} |\psi_k\rangle = \sum_{m,n}^{} \gamma_{k}
\alpha^{n}_{k} e^{ik(m+\delta_n)} |m,n\rangle,
\end{equation}
%%%%%%%%%%%%%%%%%%%%%%%%%%%%%%%%%%%%%%%%%%%%%%%%%%%%%%%%%%%%%%%%%%%%%%%%%%%%%%%
in which $k$ is the wave number, the constant value of $\delta_n$ is
$0~(0.5)$ for even~(odd) $n$,  $\alpha_k$ is given by
\begin{equation}
\label{EQ04} \alpha_k = -2 (t_1/t_2) \cos(k/2),
\end{equation}
and $\gamma_{k}$, a normalization coefficient satisfying $\langle
\psi_{k}|\psi_{k'}\rangle = \delta(k-k')$, reads
%%%%%%%%%%%%%%%%%%%%%%%%%%%%%%%%%%%%%%%%%%%%%%%%%%%%%%%%%%%%%%%%%%%%%%%%%%%%%%%
\begin{equation}
\label{EQ05} \gamma_k^2={1\over 2\pi}( 1-\alpha_k^2).
\end{equation}
%%%%%%%%%%%%%%%%%%%%%%%%%%%%%%%%%%%%%%%%%%%%%%%%%%%%%%%%%%%%%%%%%%%%%%%%%%%%%%%
Since $-2t_1<t_2$, the partial wave amplitudes appearing in the edge
wave states, $\alpha_k^{2n}$, are always less than unit and they
tend to vanish with increasing $n$. This means that the edge states
are localized at the sites of the edge chain,  and they decay to
zero by penetrating from the edge into the inside sites.
Furthermore, for the boundary values of $k$, $k=\pm\pi$, the wave
stats are zero unless $n=0$. In other words, for $k=\pm\pi$ the edge
states are strongly localized on the side zigzag chain.\par
%%%%%%%%%%%%%%%%%%%%%%%%%%%%%%%%%%%%%%%%%%%%%%%%%%%%%%%%%%%%%%%%%%%%%%%%%%%%%%%
Considering the $t_4$ term as a perturbation and applying a
first-order approximation the wave state remains unperturbed, but
the energy eigenvalue now depends on $k$ and reads
\begin{equation}
\label{EQ06} E_k=(-4t_1t_4/ t_2) - 2(2 t_1 t_4/t_2) \cos k.
\end{equation}
This expression is very similar to the dispersion relation of a
 one-dimensional tight-binding atomic chain, $ E_k=E_0 - 2 \Delta
\cos k$, where $E_0$ corresponds to $-4 t_1 t_4/ t_2$ and $\Delta$
to $2 t_1 t_4/t_2$. In other words, in present approximation an
A-type edge chain is equivalent to a one-dimensional tight-binding
chain with an integral hopping of $\Delta=2 t_1 t_4/ t_2$ and an
energy shift of $E_0=4 t_1 t_4 / t_2$.
%%%%%%%%%%%%%%%%%%%%%%%%%%%%%%%%%%%%%%%%%%%%%%%%%%%%%%%%%%%%%%%%%%%%%%%%%%%%%%%
%%%%%%%%%%%%%%%%%%%%%%%%%%%%%%%%%%%%%%%%%%%%%%%%%%%%%%%%%%%%%%%%%%%%%%%%%%%%%%%
%%%%%%%%%%%%%%%%%%%%%%%%%%%%%%%%%%%%%%%%%%%%%%%%%%%%%%%%%%%%%%%%%%%%%%%%%%%%%%%
%%%%%%%%%%%%%%%%%%%%%%%%%%%%%%%%%%%%%%%%%%%%%%%%%%%%%%%%%%%%%%%%%%%%%%%%%%%%%%%
%%%%%%%%%%%%%%%%%%%%%%%%%%%%%%%%%%%%%%%%%%%%%%%%%%%%%%%%%%%%%%%%%%%%%%%%%%%%%%%
\subsection{Green's function of the edge states in zPNRs}
With the above introduction, now we can apply a Green's function
technique to calculate the transmission coefficient of the doped
lattice for the case in which the edge-state matter waves traveling
along a zigzag chain are scattered by the on-site impurities.\par
%%%%%%%%%%%%%%%%%%%%%%%%%%%%%%%%%%%%%%%%%%%%%%%%%%%%%%%%%%%%%%%%%%%%%%%%%%%%%%%
The Green's function of a defect-free system with a discrete
spectrum is defined as
%%%%%%%%%%%%%%%%%%%%%%%%%%%%%%%%%%%%%%%%%%%%%%%%%%%%%%%%%%%%%%%%%%%%%%%%%%%%%%%
\begin{equation}
\label{EQ07} \hat{G}_E = \sum_{i}^{} {|\psi_i\rangle \langle
\psi_i|\over E-E_i+i0^+},
\end{equation}
%%%%%%%%%%%%%%%%%%%%%%%%%%%%%%%%%%%%%%%%%%%%%%%%%%%%%%%%%%%%%%%%%%%%%%%%%%%%%%%
and the same for a system of a continuous spectrum reads
%%%%%%%%%%%%%%%%%%%%%%%%%%%%%%%%%%%%%%%%%%%%%%%%%%%%%%%%%%%%%%%%%%%%%%%%%%%%%%%
\begin{equation}
\label{EQ08} \hat{G}_E = \int {|\psi_k\rangle \langle \psi_k|\over
E-E_k+i0^+} dk.
\end{equation}
%%%%%%%%%%%%%%%%%%%%%%%%%%%%%%%%%%%%%%%%%%%%%%%%%%%%%%%%%%%%%%%%%%%%%%%%%%%%%%%
For phosphorene, using the quantum edge states and their
eigenenergies given respectively in Eqs.~(\ref{EQ03})
and~(\ref{EQ06}), the matrix elements of the corresponding Green's
operator are given as
%%%%%%%%%%%%%%%%%%%%%%%%%%%%%%%%%%%%%%%%%%%%%%%%%%%%%%%%%%%%%%%%%%%%%%%%%%%%%%%
\begin{equation}
\label{EQ09}
\begin{split}
G_E (m,n;m',n') & = \langle m,n| \hat{G} | m',n' \rangle\\
&=\int_{-\pi}^{\pi} {\gamma^2_k \alpha^{n+n'}_k
e^{i(m'-m+\delta_n-\delta_{n'})}\over E-E_0 + 2 \Delta \cos k + i
0^+} dk.
\end{split}
\end{equation}
%%%%%%%%%%%%%%%%%%%%%%%%%%%%%%%%%%%%%%%%%%%%%%%%%%%%%%%%%%%%%%%%%%%%%%%%%%%%%%%
The appeared integral can be performed analytically to obtain a
closed-form expression for each of the matrix elements. In the
following discussions, we need a number of these matrix elements
which are listed as follows;
%%%%%%%%%%%%%%%%%%%%%%%%%%%%%%%%%%%%%%%%%%%%%%%%%%%%%%%%%%%%%%%%%%%%%%%%%%%%%%%
\begin{equation}
\label{EQ10}
\begin{split}
G_E(0,0;0,0) & = G_E(m,0;m,0) \\
& = {\gamma^2_{k_0}\over
2i\Delta\sin{k_0}}-{\big({2t_1/t_2}\big)^2\over 4\Delta},
\end{split}
\end{equation}
%%%%%%%%%%%%%%%%%%%%%%%%%%%%%%%%%%%%%%%%%%%%%%%%%%%%%%%%%%%%%%%%%%%%%%%%%%%%%%%
and
%%%%%%%%%%%%%%%%%%%%%%%%%%%%%%%%%%%%%%%%%%%%%%%%%%%%%%%%%%%%%%%%%%%%%%%%%%%%%%%
\begin{equation}
\label{EQ11} G_E(0,0;m,0)={\gamma_{k_0}^2 e^{ik_0m}\over
2i\Delta\sin k_0},
\end{equation}
%%%%%%%%%%%%%%%%%%%%%%%%%%%%%%%%%%%%%%%%%%%%%%%%%%%%%%%%%%%%%%%%%%%%%%%%%%%%%%%
where $k_0$ is the pole of the integrand in Eq~(\ref{EQ09}) with
$\cos k_0 = (E_0-E)/2\Delta$.\par
%%%%%%%%%%%%%%%%%%%%%%%%%%%%%%%%%%%%%%%%%%%%%%%%%%%%%%%%%%%%%%%%%%%%%%%%%%%%%%%
Also, for certain cases in which $|m-m'|\gg1$, it is easy to show
%%%%%%%%%%%%%%%%%%%%%%%%%%%%%%%%%%%%%%%%%%%%%%%%%%%%%%%%%%%%%%%%%%%%%%%%%%%%%%%
\begin{equation}
\label{EQ12} G_E(m,n;m',n') = {\gamma_{k_0}^2\alpha^{n+n'}_{k_0}
e^{i(m-m'+\delta_n-\delta_{n'}) k_0}\over 2i\Delta\sin{k_0}}.
\end{equation}\par
%%%%%%%%%%%%%%%%%%%%%%%%%%%%%%%%%%%%%%%%%%%%%%%%%%%%%%%%%%%%%%%%%%%%%%%%%%%%%%%
Furthermore,  for two sites on an armchair chain with even $n$ and
$n'$, since $m=m'$ and $\delta_n=\delta_{n'}=0$, the corresponding
matrix element of $\hat{G}$ reduces to
%%%%%%%%%%%%%%%%%%%%%%%%%%%%%%%%%%%%%%%%%%%%%%%%%%%%%%%%%%%%%%%%%%%%%%%%%%%%%%%
\begin{equation}
\label{EQ13} G_E(m,n;m,n') = \int_{-\pi}^{\pi} {\gamma^2_k
\alpha^{n+n'}_k \over E-E_0+2\Delta \cos k+ i 0^+} dk
\end{equation}
%%%%%%%%%%%%%%%%%%%%%%%%%%%%%%%%%%%%%%%%%%%%%%%%%%%%%%%%%%%%%%%%%%%%%%%%%%%%%%%
Inserting $\alpha_k$ and $\gamma_k$ from Eqs.~(\ref{EQ04})
and~(\ref{EQ05}) into the above expression, it can be separated to
two integrals. These integrals can be also evaluated analtically
using the following identity integral;
%%%%%%%%%%%%%%%%%%%%%%%%%%%%%%%%%%%%%%%%%%%%%%%%%%%%%%%%%%%%%%%%%%%%%%%%%%%%%%%
\begin{equation}
\label{EQ14}
\begin{split}
\int_{-\pi}^{+\pi}  {\cos^{2\ell} (k/2) \over E-E_0+2\Delta \cos k+
i 0^+}  dk =  {1\over 2^\ell} \sum_{j=0}^\ell {\ell \choose j} {1
\over 2^j}\\ \times  \Big[ \sum_{j'=0}^j {j \choose j'}\Big\{
\cos[(j-2j') k_0] + i \sin [|j-2j'|k_0] \Big\}\Big],
\end{split}
\end{equation}
%%%%%%%%%%%%%%%%%%%%%%%%%%%%%%%%%%%%%%%%%%%%%%%%%%%%%%%%%%%%%%%%%%%%%%%%%%%%%%%
where $\ell$ is an integer.\par
%%%%%%%%%%%%%%%%%%%%%%%%%%%%%%%%%%%%%%%%%%%%%%%%%%%%%%%%%%%%%%%%%%%%%%%%%%%%%%%
Using the obtained analytical expressions for the matrix elements of
the edge-state Green's operator and applying the general theory of
the potential scattering, we can calculate the transmission
coefficient of the system in presence of the on-site impurities
and/or vacancies. This issue will be followed in the next
section.\par
%%%%%%%%%%%%%%%%%%%%%%%%%%%%%%%%%%%%%%%%%%%%%%%%%%%%%%%%%%%%%%%%%%%%%%%%%%%%%%%
%%%%%%%%%%%%%%%%%%%%%%%%%%%%%%%%%%%%%%%%%%%%%%%%%%%%%%%%%%%%%%%%%%%%%%%%%%%%%%%
%%%%%%%%%%%%%%%%%%%%%%%%%%%%%%%%%%%%%%%%%%%%%%%%%%%%%%%%%%%%%%%%%%%%%%%%%%%%%%%
%%%%%%%%%%%%%%%%%%%%%%%%%%%%%%%%%%%%%%%%%%%%%%%%%%%%%%%%%%%%%%%%%%%%%%%%%%%%%%%
%%%%%%%%%%%%%%%%%%%%%%%%%%%%%%%%%%%%%%%%%%%%%%%%%%%%%%%%%%%%%%%%%%%%%%%%%%%%%%%
\section{Electrical conductance of the zigzag-edged phosphorene nanoribbons}
In this section, we present the results of the transport
calculations in the pure and the defective zPNRs. Our aim of this
study is the exploring of the influence of the defect type and its
position on the transport properties of the considered system. To
perform the calculations, we begin with the Lippmann-Schwinger
equation in the potential scattering theory
%%%%%%%%%%%%%%%%%%%%%%%%%%%%%%%%%%%%%%%%%%%%%%%%%%%%%%%%%%%%%%%%%%%%%%%%%%%%%%%
\begin{equation}
\label{EQ15}\begin{split} |\psi_{out} \rangle & = |\psi_{in}
\rangle+\hat{G}_E\hat{V}|\psi_{out} \rangle\\ & =|\psi_{in}
\rangle+\hat{G}_E\hat{T}|\psi_{in} \rangle,\end{split}
\end{equation}
%%%%%%%%%%%%%%%%%%%%%%%%%%%%%%%%%%%%%%%%%%%%%%%%%%%%%%%%%%%%%%%%%%%%%%%%%%%%%%%
where $|\psi_{in} \rangle$ and $|\psi_{out} \rangle$ are the quantum
states of the incoming and outgoing charge carriers, $\hat{G}_E$ is
the Green operator for the unperturbed system, $\hat{V}$ is the
scattering potential and $\hat{T}$ is the transition matrix
corresponding to the scattering potential $\hat{V}$ with
$\hat{V}|\psi_{out} \rangle=\hat{T}|\psi_{in} \rangle$. An attempt
to solve the Lippmann-Schwinger equation by successive iterations
leads to
%%%%%%%%%%%%%%%%%%%%%%%%%%%%%%%%%%%%%%%%%%%%%%%%%%%%%%%%%%%%%%%%%%%%%%%%%%%%%%%
\begin{equation}
\label{EQ16} \hat{T}=\hat{V} ( 1 + \hat{G}_E \hat{V} + \hat{G}_E
\hat{V}\hat{G}_E \hat{V} +\cdots).
\end{equation}
%%%%%%%%%%%%%%%%%%%%%%%%%%%%%%%%%%%%%%%%%%%%%%%%%%%%%%%%%%%%%%%%%%%%%%%%%%%%%%%
%%%%%%%%%%%%%%%%%%%%%%%%%%%%%%%%%%%%%%%%%%%%%%%%%%%%%%%%%%%%%%%%%%%%%%%%%%%%%%%
Now, we assume that two on-site  defects are localized at positions
$(m_1,n_1)$ and $(m_2,n_2)$, and consider the edge-state charge
carriers traveling along the side zigzag chain. These traveling
matter waves will be scattered by the defects. Consequently, the
incoming wave in the Lippmann-Schwinger equation is the edge-state
wave given in Eq.~(\ref{EQ03}).  The, scattering potential due to
the presence of the defects is given in its explicit form as
%%%%%%%%%%%%%%%%%%%%%%%%%%%%%%%%%%%%%%%%%%%%%%%%%%%%%%%%%%%%%%%%%%%%%%%%%%%%%%%
\begin{equation}
\label{EQ17} \hat{V}=V_1 |m_1,n_1\rangle \langle m_1,n_1| +
V_2|m_2,n_2\rangle \langle m_2,n_2|,
\end{equation}
%%%%%%%%%%%%%%%%%%%%%%%%%%%%%%%%%%%%%%%%%%%%%%%%%%%%%%%%%%%%%%%%%%%%%%%%%%%%%%%
where $V_1$ and $V_2$ are the potential strengthes due to the
presence of the defects. With this form of the scattering potential
the transition operator $\hat T$ can be written in its matrix form
as
%%%%%%%%%%%%%%%%%%%%%%%%%%%%%%%%%%%%%%%%%%%%%%%%%%%%%%%%%%%%%%%%%%%%%%%%%%%%%%%
\begin{equation}
\label{EQ18} \hat{T}=[\hat{V}]+
(1+[\hat{G}_E][\hat{V}]+[\hat{G}_E][\hat{V}][\hat{G}_E][\hat{V}]+\cdots),
\end{equation}
%%%%%%%%%%%%%%%%%%%%%%%%%%%%%%%%%%%%%%%%%%%%%%%%%%%%%%%%%%%%%%%%%%%%%%%%%%%%%%%
where
%%%%%%%%%%%%%%%%%%%%%%%%%%%%%%%%%%%%%%%%%%%%%%%%%%%%%%%%%%%%%%%%%%%%%%%%%%%%%%%
\begin{equation}
\label{EQ19} [\hat{V}]=\left(\begin{array}{cc}
V_1 & 0 \\
0 & V_2
\end{array}\right),\ \rm{and}\ [\hat{G}_E]= \left(\begin{array}{cc}
G_{11} & G_{12} \\
G_{21} & G_{22}
\end{array}\right).
\end{equation}
%%%%%%%%%%%%%%%%%%%%%%%%%%%%%%%%%%%%%%%%%%%%%%%%%%%%%%%%%%%%%%%%%%%%%%%%%%%%%%%
It is obvious that the matrix elements of $[\hat{G}_E]$ are known as
%%%%%%%%%%%%%%%%%%%%%%%%%%%%%%%%%%%%%%%%%%%%%%%%%%%%%%%%%%%%%%%%%%%%%%%%%%%%%%%
\begin{equation}
\label{EQ20}
\begin{split}
G_{11}=G_E(m_1,n_1;m_2,n_2),\cr
G_{12}=G_E(m_1,n_1;m_2,n_2),\cr
G_{21}=G_E(m_2,n_2;m_1,n_1),\cr
G_{22}=G_E(m_2,n_2;m_2,n_2).\end{split}
\end{equation}
%%%%%%%%%%%%%%%%%%%%%%%%%%%%%%%%%%%%%%%%%%%%%%%%%%%%%%%%%%%%%%%%%%%%%%%%%%%%%%%
Since, the matrix forms of the scattering potential and the
free-defect Green's operator are exactly known, a closed-form
expression can be evaluated for the transition matrix using the fact
that
%%%%%%%%%%%%%%%%%%%%%%%%%%%%%%%%%%%%%%%%%%%%%%%%%%%%%%%%%%%%%%%%%%%%%%%%%%%%%%%
\begin{equation}
\label{EQ21} \hat{T} = [\hat{V}] \left(1- [\hat{G}_E]
[\hat{V}]\right)^{-1}.
\end{equation}
%%%%%%%%%%%%%%%%%%%%%%%%%%%%%%%%%%%%%%%%%%%%%%%%%%%%%%%%%%%%%%%%%%%%%%%%%%%%%%%
%%%%%%%%%%%%%%%%%%%%%%%%%%%%%%%%%%%%%%%%%%%%%%%%%%%%%%%%%%%%%%%%%%%%%%%%%%%%%%%
Substituting the edge-state wave function as the incoming wave, and
the matrix forms of $\hat{G}_E$ and~$\hat{T}$ into Eq.~(\ref{EQ15}),
the outgoing wave function can be calculated. To this end, we
introduce the quantum state $|\chi\rangle$ as
%%%%%%%%%%%%%%%%%%%%%%%%%%%%%%%%%%%%%%%%%%%%%%%%%%%%%%%%%%%%%%%%%%%%%%%%%%%%%%%
\begin{equation}
\label{EQ22} |\chi\rangle= \hat{T}| \psi_{in}\rangle = \alpha |
n_1,m_1 \rangle + \beta |m_2,n_2\rangle,
\end{equation}
%%%%%%%%%%%%%%%%%%%%%%%%%%%%%%%%%%%%%%%%%%%%%%%%%%%%%%%%%%%%%%%%%%%%%%%%%%%%%%%
where
%%%%%%%%%%%%%%%%%%%%%%%%%%%%%%%%%%%%%%%%%%%%%%%%%%%%%%%%%%%%%%%%%%%%%%%%%%%%%%%
\begin{equation}
\begin{split} \alpha & = T_{11} \langle n_1,m_1 | \psi_{in} \rangle +
T_{12}  \langle n_2,m_2 | \psi_{in} \rangle ,\\
\beta & =  T_{21} \langle n_1,m_1 | \psi_{in} \rangle + T_{22}
\langle n_2,m_2 | \psi_{in} \rangle, \end{split}\label{EQ23}
\end{equation}
%%%%%%%%%%%%%%%%%%%%%%%%%%%%%%%%%%%%%%%%%%%%%%%%%%%%%%%%%%%%%%%%%%%%%%%%%%%%%%%
in which $T_{11}$, $T_{12}$, $T_{21}$ and $T_{22}$ are the
calculated matrix elements of $\hat{T}$ for the specified case.
Having $|\chi\rangle$, we find out
%%%%%%%%%%%%%%%%%%%%%%%%%%%%%%%%%%%%%%%%%%%%%%%%%%%%%%%%%%%%%%%%%%%%%%%%%%%%%%%
\begin{equation}
\label{EQ24} | \psi_{out}\rangle = | \psi_{in}\rangle +\hat{G}_E
|\chi \rangle.
\end{equation}
%%%%%%%%%%%%%%%%%%%%%%%%%%%%%%%%%%%%%%%%%%%%%%%%%%%%%%%%%%%%%%%%%%%%%%%%%%%%%%%
%%%%%%%%%%%%%%%%%%%%%%%%%%%%%%%%%%%%%%%%%%%%%%%%%%%%%%%%%%%%%%%%%%%%%%%%%%%%%%%
Using the obtained result for the outgoing wave, both the
transmitted and the reflected parts of the final sate are derivable.
Having these parts of the outgoing wave, the reflection and
transmission coefficients can be evaluated. To evaluate these
coefficients, for an arbitrary site of $(m',n')$ it is seen that
%%%%%%%%%%%%%%%%%%%%%%%%%%%%%%%%%%%%%%%%%%%%%%%%%%%%%%%%%%%%%%%%%%%%%%%%%%%%%%%
\begin{equation}
\label{EQ25} \begin{split} \langle m' , n' | \hat{G}_E |\chi \rangle
= \alpha G_E(m',n'; m_1,n_1) \\ +  \beta G_E(m' ,
n';m_2,n_2),\end{split}
\end{equation}
%%%%%%%%%%%%%%%%%%%%%%%%%%%%%%%%%%%%%%%%%%%%%%%%%%%%%%%%%%%%%%%%%%%%%%%%%%%%%%%
Since the transmission coefficient is more of interest, we examine
the transmitted part of the final wave state at an arbitrary
position of $(m',n')$ with $m' \gg m_2$. In this case, using the
form of the Green's operator it is easy to show that the above
expression is proportional to $\langle m' , n' |\psi_k\rangle$.
Consequently, we obtain the transmission amplitude as
%%%%%%%%%%%%%%%%%%%%%%%%%%%%%%%%%%%%%%%%%%%%%%%%%%%%%%%%%%%%%%%%%%%%%%%%%%%%%%%
\begin{equation}
\label{EQ26} \tau = 1 + {\alpha G_E(m',n'; m_1,n_1) + \beta G_E(m' ,
n'; m_2, n_2)\over \langle m' , n' |\psi_k\rangle},
\end{equation}
%%%%%%%%%%%%%%%%%%%%%%%%%%%%%%%%%%%%%%%%%%%%%%%%%%%%%%%%%%%%%%%%%%%%%%%%%%%%%%%
in which it is obviously known that $m'\gg m_2$. Finally the
transmission coefficient is obtained using ${\mathcal
T}(E)=|\tau|^2$.\par
%%%%%%%%%%%%%%%%%%%%%%%%%%%%%%%%%%%%%%%%%%%%%%%%%%%%%%%%%%%%%%%%%%%%%%%%%%%%%%%
%%%%%%%%%%%%%%%%%%%%%%%%%%%%%%%%%%%%%%%%%%%%%%%%%%%%%%%%%%%%%%%%%%%%%%%%%%%%%%%
%%%%%%%%%%%%%%%%%%%%%%%%%%%%%%%%%%%%%%%%%%%%%%%%%%%%%%%%%%%%%%%%%%%%%%%%%%%%%%%
\begin{figure}[t]
\centering
\includegraphics[scale=0.6]{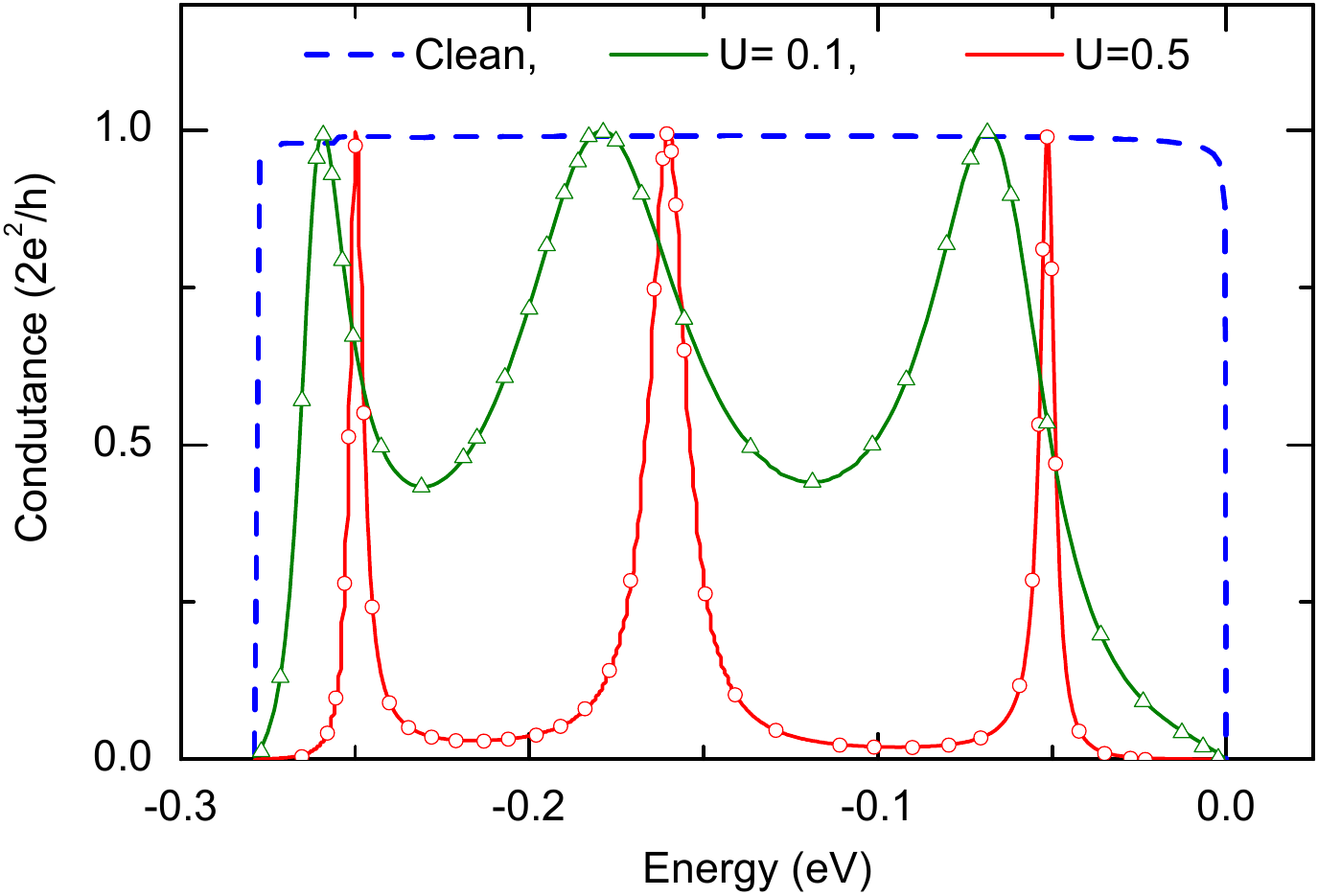}
\caption{(Color online) Conductance of a defective zPNR with two
atomic impurities fixed at $(0,0)$ and $(4,0)$. Two values of the
interaction potential are considered, $U=0.1~eV$ and $U=0.5~eV$. The
lines are for the present analytical calculations and the symbols
are for their corresponding numerical simulations.} \label{Fig03}
\end{figure}
%%%%%%%%%%%%%%%%%%%%%%%%%%%%%%%%%%%%%%%%%%%%%%%%%%%%%%%%%%%%%%%%%%%%%%%%%%%%%%%
%%%%%%%%%%%%%%%%%%%%%%%%%%%%%%%%%%%%%%%%%%%%%%%%%%%%%%%%%%%%%%%%%%%%%%%%%%%%%%%
%%%%%%%%%%%%%%%%%%%%%%%%%%%%%%%%%%%%%%%%%%%%%%%%%%%%%%%%%%%%%%%%%%%%%%%%%%%%%%%
In the following discussion, we consider two on-site impurities
and/or vacancies in the nanoribbon. As is seen in Fig.~\ref{Fig01},
one of the impurities/vacances is located at one of the sites of the
zigzag side of the nanoribbon. This site is labeled as
$(m_1,n_1)=(0,0)$. For the second impurity/vacancy, we consider two
different positions. In the first case, the impurity/vacancy is also
located at one of the sites of the side zigzag chain. So, this site
is labeled as $(m_2,n_2)=(m,0)$. For the second case, we consider
the second defect located at the same armchair chain as the first
one, but the related zigzag chain is labeled by an even integer. So,
the defected site in this case is labeled as $(m_2,n_2)=(0,n)$ where
$n$ is even.\par
%%%%%%%%%%%%%%%%%%%%%%%%%%%%%%%%%%%%%%%%%%%%%%%%%%%%%%%%%%%%%%%%%%%%%%%%%%%%%%%
Also, for simplicity, we assume that the potential strengths $V_1$
and $V_2$ are equal, so we put $V_1=V_2=U$, and we notice that for
two vacancy defects, $U$ tends to infinity ($U\to\infty$).\par
%%%%%%%%%%%%%%%%%%%%%%%%%%%%%%%%%%%%%%%%%%%%%%%%%%%%%%%%%%%%%%%%%%%%%%%%%%%%%%%
%%%%%%%%%%%%%%%%%%%%%%%%%%%%%%%%%%%%%%%%%%%%%%%%%%%%%%%%%%%%%%%%%%%%%%%%%%%%%%%
\section{Results and discussion\label{sec03}}
In all the following discussions, we assumed that the first
impurity/vacancy is fixed at $(m_1,n_1)=(0,0)$ on the edge zigzag
chain.\par
%%%%%%%%%%%%%%%%%%%%%%%%%%%%%%%%%%%%%%%%%%%%%%%%%%%%%%%%%%%%%%%%%%%%%%%%%%%%%%%
%%%%%%%%%%%%%%%%%%%%%%%%%%%%%%%%%%%%%%%%%%%%%%%%%%%%%%%%%%%%%%%%%%%%%%%%%%%%%%%
Figure~\ref{Fig03} shows the conductance as a function of the
electron energy for the case in which the second impurity is located
on the edge zigzag chain at $(m_2=m, n_2=0)$. So, both defects are
along the $x$ axis. The graphs are plotted for $m=4$ and for two
values of potential strength, $U=0.1$ and $U=0.5$. As is seen the
conductance presents a resonant behavior which is expected for
successive reflection and transmissions of the matter waves from the
impurities and the constructive or destructive interference of the
transmitted parts. The number of the observed resonant peaks are
equal to the number of the atomic sites between the impurities. For
the weaker interaction, $U=0.1$, the peaks are not very sharp, but
with increasing the interaction strength the peaks become sharper
and shift toward the larger energies. The numerical simulations
performed using the approach followed in Ref.~\cite{Amini2018} are
also depicted in this figure. As is seen, our present analytical
results are in excellent agreement with the numerical
simulations.\par
%%%%%%%%%%%%%%%%%%%%%%%%%%%%%%%%%%%%%%%%%%%%%%%%%%%%%%%%%%%%%%%%%%%%%%%%%%%%%%%
%%%%%%%%%%%%%%%%%%%%%%%%%%%%%%%%%%%%%%%%%%%%%%%%%%%%%%%%%%%%%%%%%%%%%%%%%%%%%%%
\begin{figure}[t]
\centering
\includegraphics[scale=0.6]{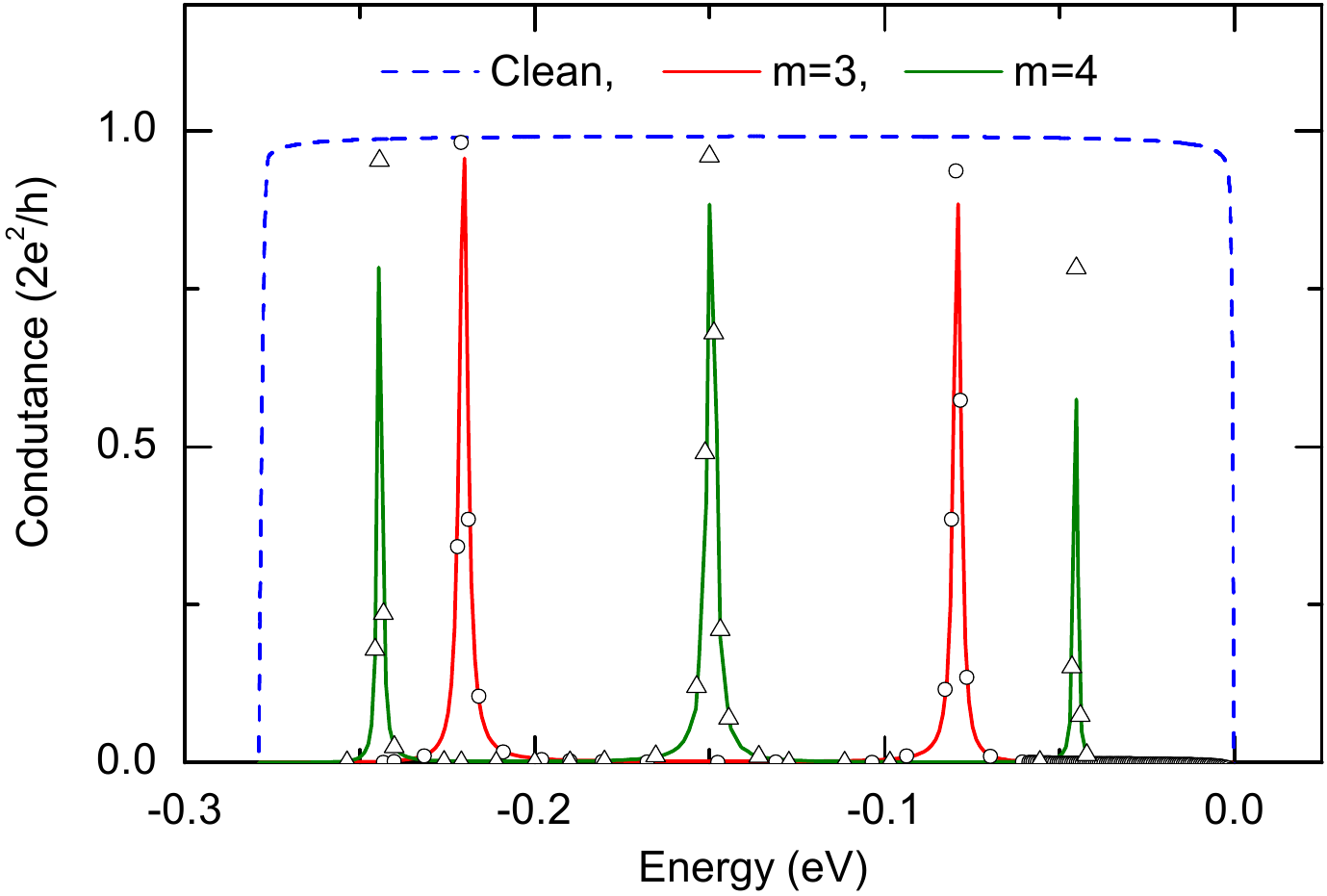}
\caption{(Color online) Same as figure~\ref{Fig03} but for two
vacancies on the edge zigzag chain. The first vacancy is fixed at
$(0,0)$, but for the second vacancy two positions are considered:
$(3,0)$ and $(4,0)$.} \label{Fig04}
\end{figure}
%%%%%%%%%%%%%%%%%%%%%%%%%%%%%%%%%%%%%%%%%%%%%%%%%%%%%%%%%%%%%%%%%%%%%%%%%%%%%%%
%%%%%%%%%%%%%%%%%%%%%%%%%%%%%%%%%%%%%%%%%%%%%%%%%%%%%%%%%%%%%%%%%%%%%%%%%%%%%%%
A similar situation for two vacancies on the edge zigzag chain is
demonstrated in Fig.~\ref{Fig04}. For this case, $U$ tends to
infinity. The resonant peaks which are also seen in this case are
extremely sharp and the number of peaks are equal to $m-1$.
Figures~\ref{Fig03} and~\ref{Fig04} show that the behavior of the
considered system is similar to what happens for a one-dimensional
tight binding chain. For a one-dimensional chain the vacancy limit
is meaningless, because this is equivalent to the chain being cut
off. But for the present quasi one-dimensional system this limit is
completely meaningful.\par
%%%%%%%%%%%%%%%%%%%%%%%%%%%%%%%%%%%%%%%%%%%%%%%%%%%%%%%%%%%%%%%%%%%%%%%%%%%%%%%
%%%%%%%%%%%%%%%%%%%%%%%%%%%%%%%%%%%%%%%%%%%%%%%%%%%%%%%%%%%%%%%%%%%%%%%%%%%%%%%
\begin{figure}[t]
\centering
\includegraphics[scale=0.6]{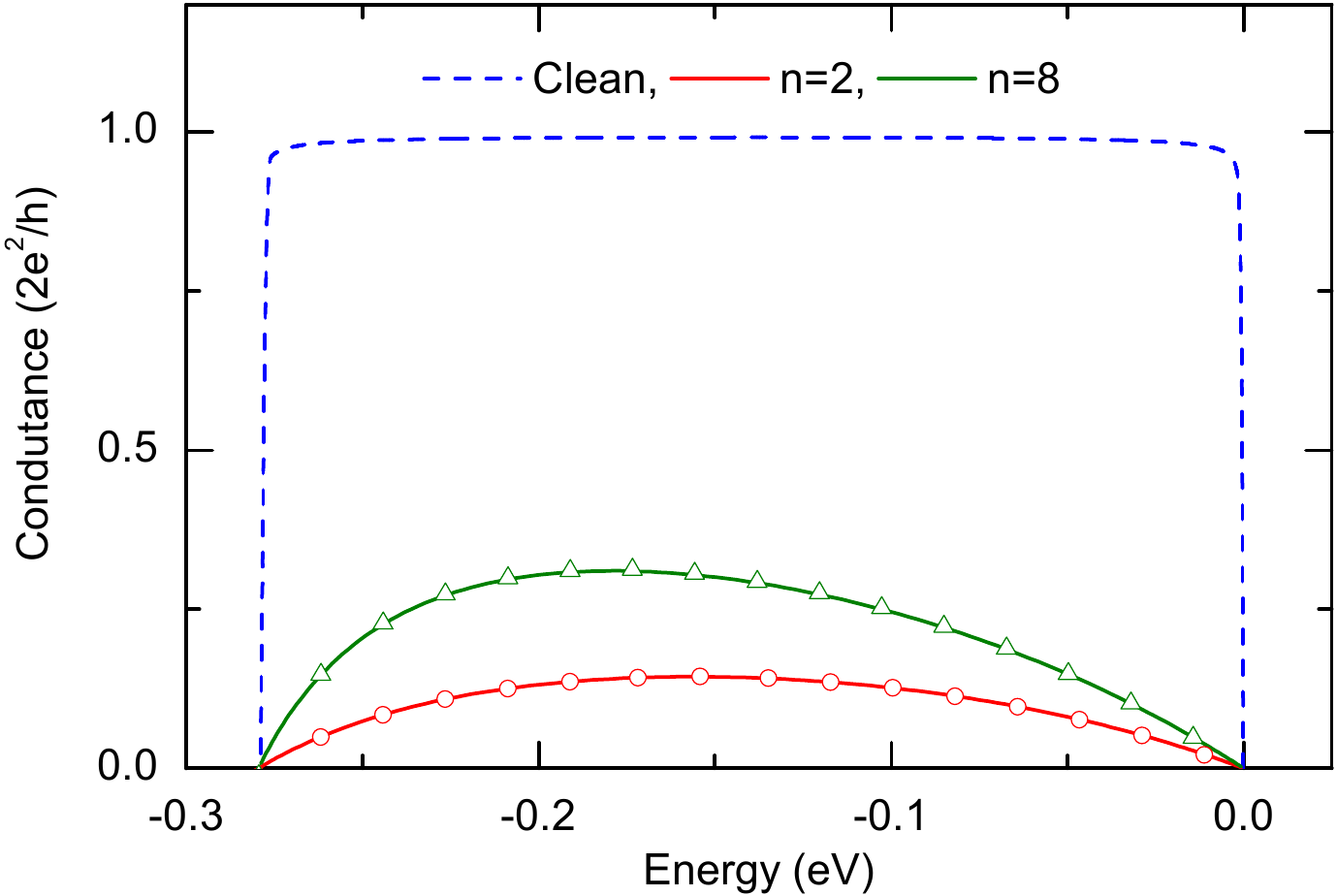}
\caption{(Color online) Electrical conductivity as a function of
energy for a case in which two atomic impurities are located in $y$
direction. The first impurity is on the edge zigzag chain and the
second is inside the bulk. Two different positions for the second
impurity are considered $(0,2)$ and $(0,8)$. } \label{Fig05}
\end{figure}
%%%%%%%%%%%%%%%%%%%%%%%%%%%%%%%%%%%%%%%%%%%%%%%%%%%%%%%%%%%%%%%%%%%%%%%%%%%%%%%
%%%%%%%%%%%%%%%%%%%%%%%%%%%%%%%%%%%%%%%%%%%%%%%%%%%%%%%%%%%%%%%%%%%%%%%%%%%%%%%
Figure~\ref{Fig05} shows the conductivity as a function of energy
for two impurities located along the $y$ axis. The first impurity is
on the edge zigzag chain and the second is inside the bulk. The
defected sites are denoted as $(0,0)$ and $(0,n)$, where $n$ is an
even integer. As is seen, in this case the conductivity is in
general weak. Whatever the second impurity penetrates more into the
bulk its effect will be less. So that, for the large values of $n$
the behavior of the system is closer to a system with a single
impurity placed on the edge zigzag chain.\par
%%%%%%%%%%%%%%%%%%%%%%%%%%%%%%%%%%%%%%%%%%%%%%%%%%%%%%%%%%%%%%%%%%%%%%%%%%%%%%%
%%%%%%%%%%%%%%%%%%%%%%%%%%%%%%%%%%%%%%%%%%%%%%%%%%%%%%%%%%%%%%%%%%%%%%%%%%%%%%%
\begin{figure}[t]
\centering
\includegraphics[scale=0.6]{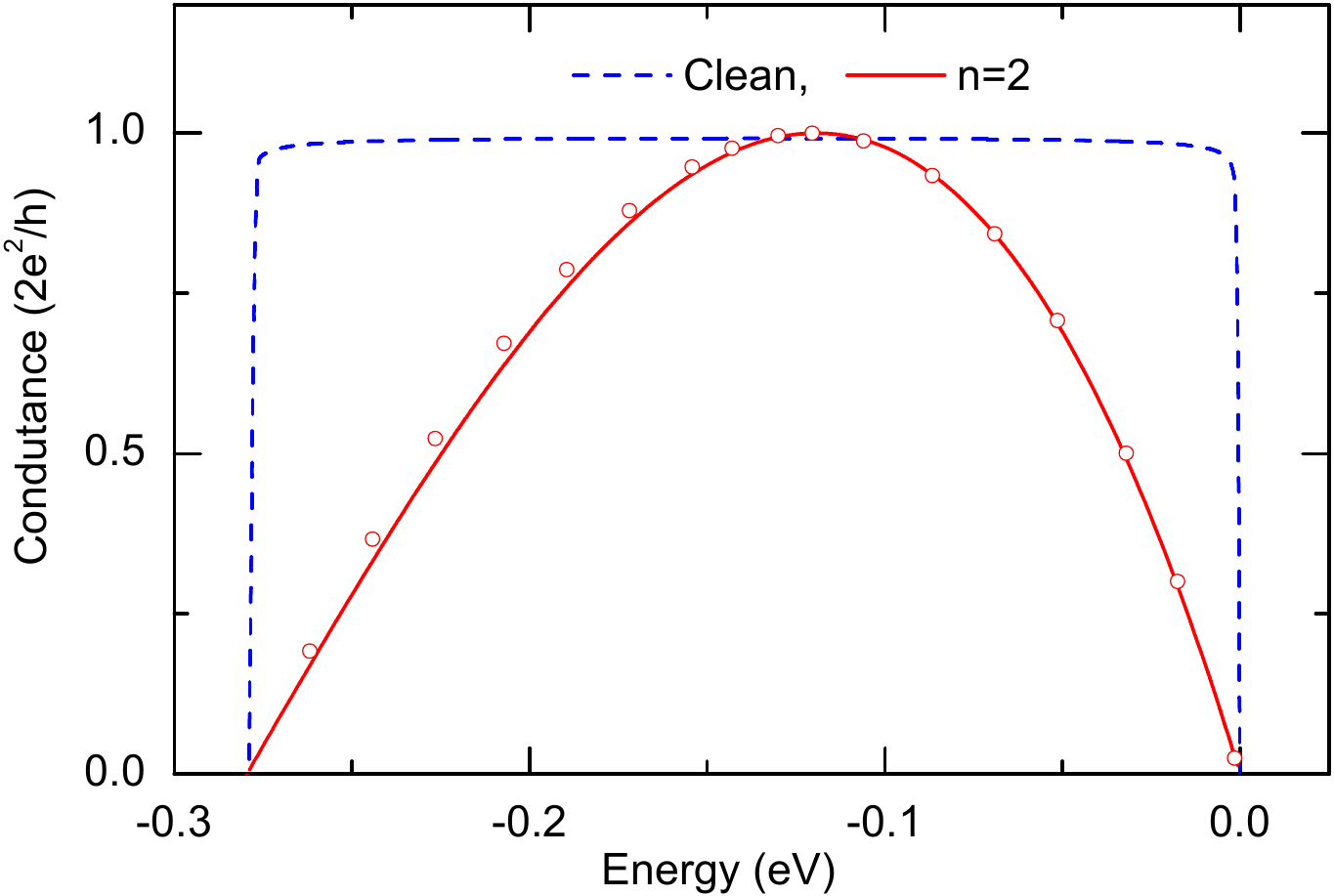}
\caption{(Color online) Same as figure~\ref{Fig05} but for two
vacancies on $y$ axis. The second vacancy is created at position of
$(0,2)$.} \label{Fig06}
\end{figure}
%%%%%%%%%%%%%%%%%%%%%%%%%%%%%%%%%%%%%%%%%%%%%%%%%%%%%%%%%%%%%%%%%%%%%%%%%%%%%%%
%%%%%%%%%%%%%%%%%%%%%%%%%%%%%%%%%%%%%%%%%%%%%%%%%%%%%%%%%%%%%%%%%%%%%%%%%%%%%%%
An interesting situation occurs when two vacancies are created along
the $y$~axis. Figure~\ref{Fig06} shows that if, instead of two
atomic impurities, two vacancies are placed on an armchair chain
along the $y$~axis the electrical conduction increases considerably.
As is seen, in a certain range of the electron energy, the
conductivity is maximized. Since the presence of the second vacancy
not only does not reduce the conductivity but increases it
significantly, this situation is very interesting. However, the
behavior of conductivity by changing the energy in this case is
non-resonant. This amazing behavior can be explained as follows. If
we consider only the second vacancy which is located on $(0,2)$ and
calculate the unperturbed Green's operator, the result can be used
to calculate LDoS on the first vacancy site, $(0,0)$. We see that
LDoS changes strongly. If, we use the Lippmann-Schwinger equation to
study the scattering of the incoming matter waves from the first
vacancy, it seems that due to the presence of the second vacancy,
these waves turn around this vacancy. In other words, the presence
of the second vacancy reduces the penetration of the traveling waves
into the spatial region around the first vacancy. So, the second
vacancy creates a shielding around the first vacancy. In this case,
for $n=2$ the agreement between the present analytical results and
their corresponding numerical simulations is excellent, but with
increasing $n$, some discrepancies are observed between analytical
and numerical calculations. This is due to the creation of the
localized states with resonant energies in the band structure. The
details of this issue are under consideration.
%%%%%%%%%%%%%%%%%%%%%%%%%%%%%%%%%%%%%%%%%%%%%%%%%%%%%%%%%%%%%%%%%%%%%%%%%%%%%%%
%%%%%%%%%%%%%%%%%%%%%%%%%%%%%%%%%%%%%%%%%%%%%%%%%%%%%%%%%%%%%%%%%%%%%%%%%%%%%%%
\section{Summary and conclusion}\label{V}
%%%%%%%%%%%%%%%%%%%%%%%%%%%%%%%%%%%%%%%%%%%%%%%%%%%%%%%%%%%%%%%%%%%%%%%%%%%%%%%
Considering a tight-binding model and applying the general
scattering theory, the electron transport in a zPNR in presence of
two impurities/vacancies was analytically investigated. For two
impurities on the edge zigzag chain, the variations of the
conductivity with the electron energy exhibits a resonant behavior.
For a similar situation with two vacancies, a similar resonant
behavior with extremely sharp peaks has been observed. It was shown
that the agreement between the analytical calculations and the
numerical simulations in these cases is excellent. It found out that
the presence of the impurities along the $y$-axis reduces the
conductivity considerably. If the distance between the impurities
increases the behavior of the system approaches the behavior of a
system with a single impurity on the zigzag chain. Two adjacent
vacancies induce  considerable conduction in a certain range of the
electron energy. This is due to creating a shielding by the second
vacancy around the first one. For this final case, the agreement
between the analytical and numerical calculations is excellent only
for $n=2$. For larger value of $n$, although the overall behavior of
analytical calculations and simulations is the same, they have
considerable discrepancies in detail.\par
%%%%%%%%%%%%%%%%%%%%%%%%%%%%%%%%%%%%%%%%%%%%%%%%%%%%%%%%%%%%%%%%%%%%%%%%%%%%%%%
%%%%%%%%%%%%%%%%%%%%%%%%%%%%%%%%%%%%%%%%%%%%%%%%%%%%%%%%%%%%%%%%%%%%%%%%%%%%%%%
%%%%%%%%%%%%%%%%%%%%%%%%%%%%%%%%%%%%%%%%%%%%%%%%%%%%%%%%%%%%%%%%%%%%%%%%%%%%%%%
\acknowledgments 
MA acknowledges hospitality of the International Centre for Theoretical Physics~(ICTP), Trieste, Italy.
The fourth author also would like to acknowledge the
office of graduate studies at the University of Isfahan for their
support and research facilities.
%%%%%%%%%%%%%%%%%%%%%%%%%%%%%%%%%%%%%%%%%%%%%%%%%%%%%%%%%%%%%%%%%%%%%%%%%%%%%%%
%%%%%%%%%%%%%%%%%%%%%%%%%%%%%%%%%%%%%%%%%%%%%%%%%%%%%%%%%%%%%%%%%%%%%%%%%%%%%%%
%%%%%%%%%%%%%%%%%%%%%%%%%%%%%%%%%%%%%%%%%%%%%%%%%%%%%%%%%%%%%%%%%%%%%%%%%%%%%%%

%%%%%%%%%%%%%%%%%%%%%%%%%%%%%%%%%%%%%%%%%%%%%%%%%%%%%%%%%%%%%%%%%%%%%%%%%%%%%%%
\end{document}